\documentclass[sigconf]{acmart}

\usepackage{algorithm}
\usepackage{algorithmic}
\usepackage{amsmath}
\usepackage{array}
\usepackage{diagbox}
\usepackage{makecell}
\usepackage{graphicx}
\usepackage{float}
\usepackage{subfig}
\usepackage{enumitem}
\usepackage{float}
\usepackage[frozencache,cachedir=.]{minted}

\newcommand{\ours}{EICopilot}

\begin{document}
\title{EICopilot: Search and Explore Enterprise Information over Large-scale Knowledge Graphs with LLM-driven Agents}



\author{Yuhui Yun\textsuperscript{*}}
\affiliation{%
  \institution{Baidu Inc.}
  \city{Haidian District, Beijing}
  \country{China}
}

\author{Huilong Ye\textsuperscript{*}}
\affiliation{%
  \institution{Baidu Inc.}
  \city{Haidian District, Beijing}
  \country{China}}

\author{Xinru Li}
\affiliation{%
  \institution{Wuhan University}
  \city{WuHan, Hubei}
  \country{China}
}

\author{Ruojia Li}
\affiliation{%
  \institution{University of Chinese Academy of Sciences}
  \city{Haidian District, Beijing}
  \country{China}}

\author{Jingfeng Deng}
\affiliation{%
  \institution{South China University of Technology}
  \city{Guangzhou}
  \country{China}}

\author{Li Li}
\affiliation{%
  \institution{University of Macau}
  \city{Macau}
  \country{China}}
  
\author{Haoyi Xiong}
\affiliation{%
  \institution{Baidu Inc.}
  \city{Haidian District, Beijing}
  \country{China}
}
\authornote{These authors contributed equally to this research.}
\authornote{Correspondence to: Haoyi Xiong \textless haoyi.xiong.fir@ieee.org\textgreater.}


\begin{abstract}
The paper introduces \ours{}, an novel agent-based solution enhancing search and exploration of enterprise registration data within extensive online knowledge graphs like those detailing legal entities, registered capital, and major shareholders. Traditional methods necessitate text-based queries and manual subgraph explorations, often resulting in time-consuming processes. \ours{}, deployed as a chatbot via Baidu Enterprise Search, improves this landscape by utilizing Large Language Models (LLMs) to interpret natural language queries. This solution automatically generates and executes Gremlin scripts, providing efficient summaries of complex enterprise relationships. Distinct feature a data pre-processing pipeline that compiles and annotates representative queries into a vector database of examples for In-context learning (ICL), a comprehensive reasoning pipeline combining Chain-of-Thought with ICL to enhance Gremlin script generation for knowledge graph search and exploration, and a novel query masking strategy that improves intent recognition for heightened script accuracy. Empirical evaluations demonstrate the superior performance of \ours{}, including speed and accuracy, over baseline methods, with the \emph{Full Mask} variant achieving a syntax error rate reduction to as low as 10.00\% and an execution correctness of up to 82.14\%. These components collectively contribute to superior querying capabilities and summarization of intricate datasets, positioning \ours{} as a groundbreaking tool in the exploration and exploitation of large-scale knowledge graphs for enterprise information search.  

\end{abstract}




\maketitle

\section{Introduction}
Online knowledge graphs detailing vital enterprise registration data, such as legal persons, registered capital, and major shareholders, serve as a valuable resource for Internet users seeking enterprise information\cite{li2024survey}. Despite their utility, exploring these graphs can be cumbersome due to the need for intricate text-based queries and manual exploration of subgraphs, presenting significant challenges in extracting pertinent information efficiently\cite{qiu2020hierarchical}. 

\begin{figure*}[ht]
    \centering
    \subfloat[Manual enterprise search]{\includegraphics[width=0.42\textwidth]{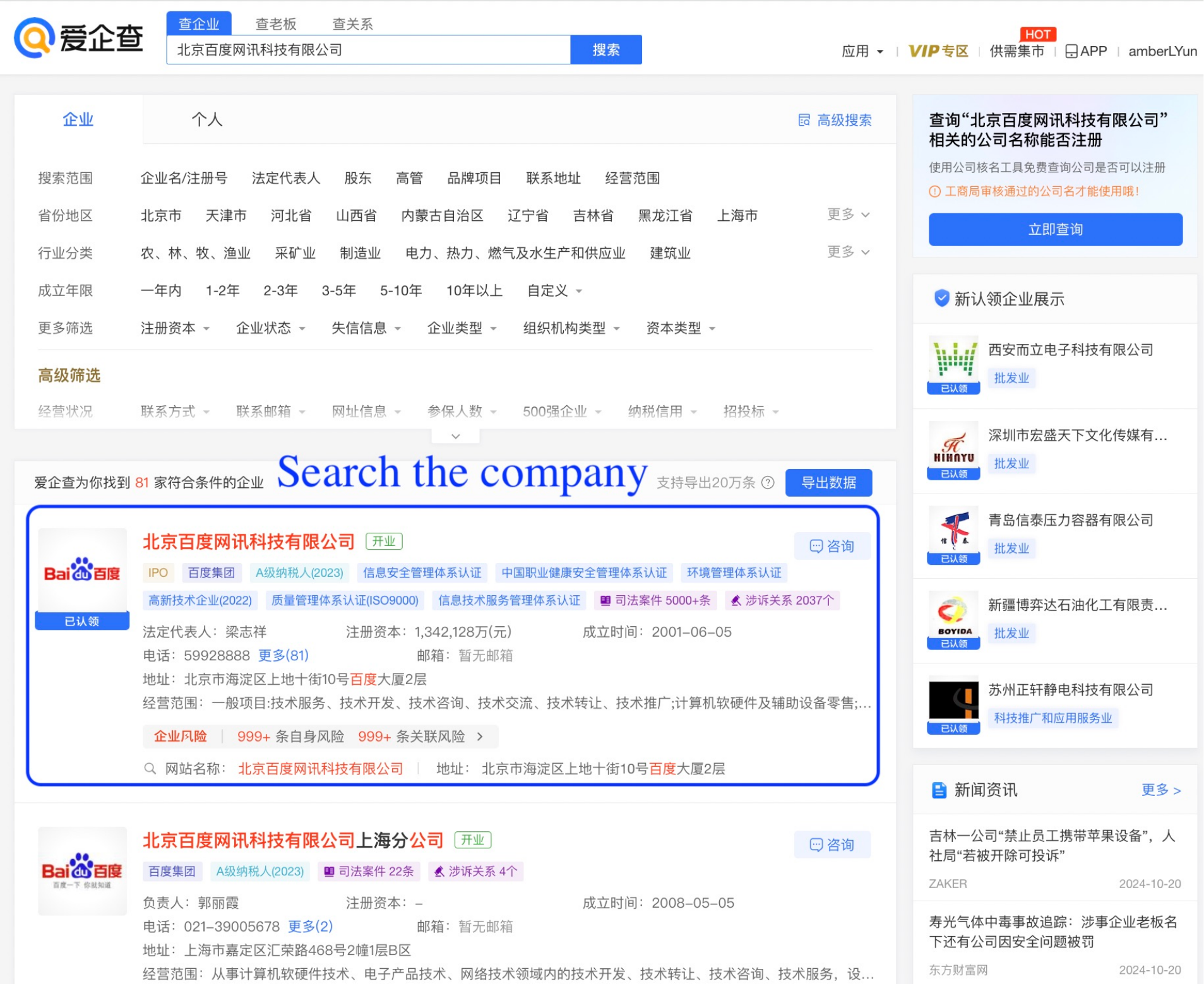}} \quad
    \subfloat[Manual shareholder exploration]{\includegraphics[width=0.42\textwidth]{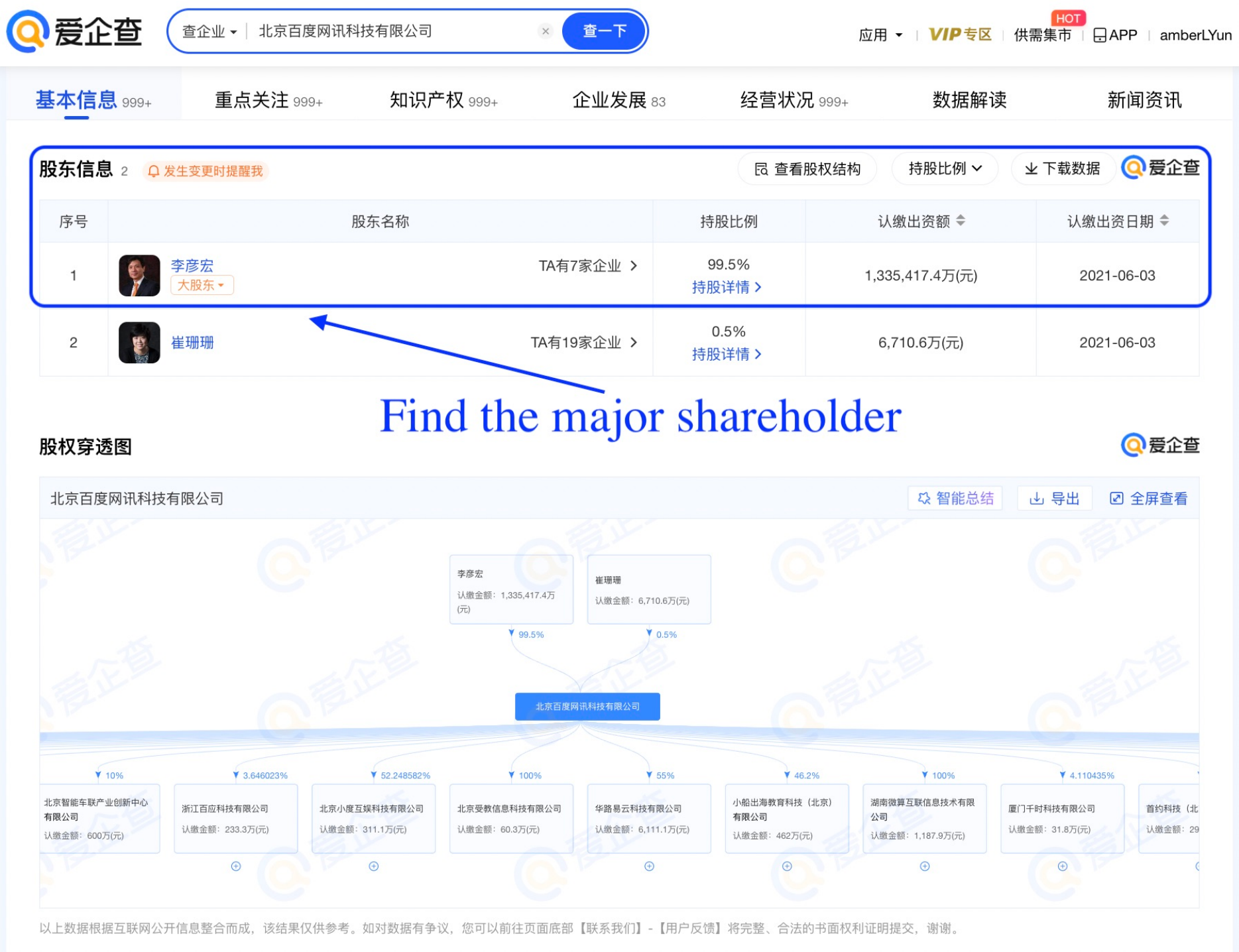}} \\ \vspace{-3mm}
    \subfloat[Manual catering enterprise search]{\includegraphics[width=0.42\textwidth]{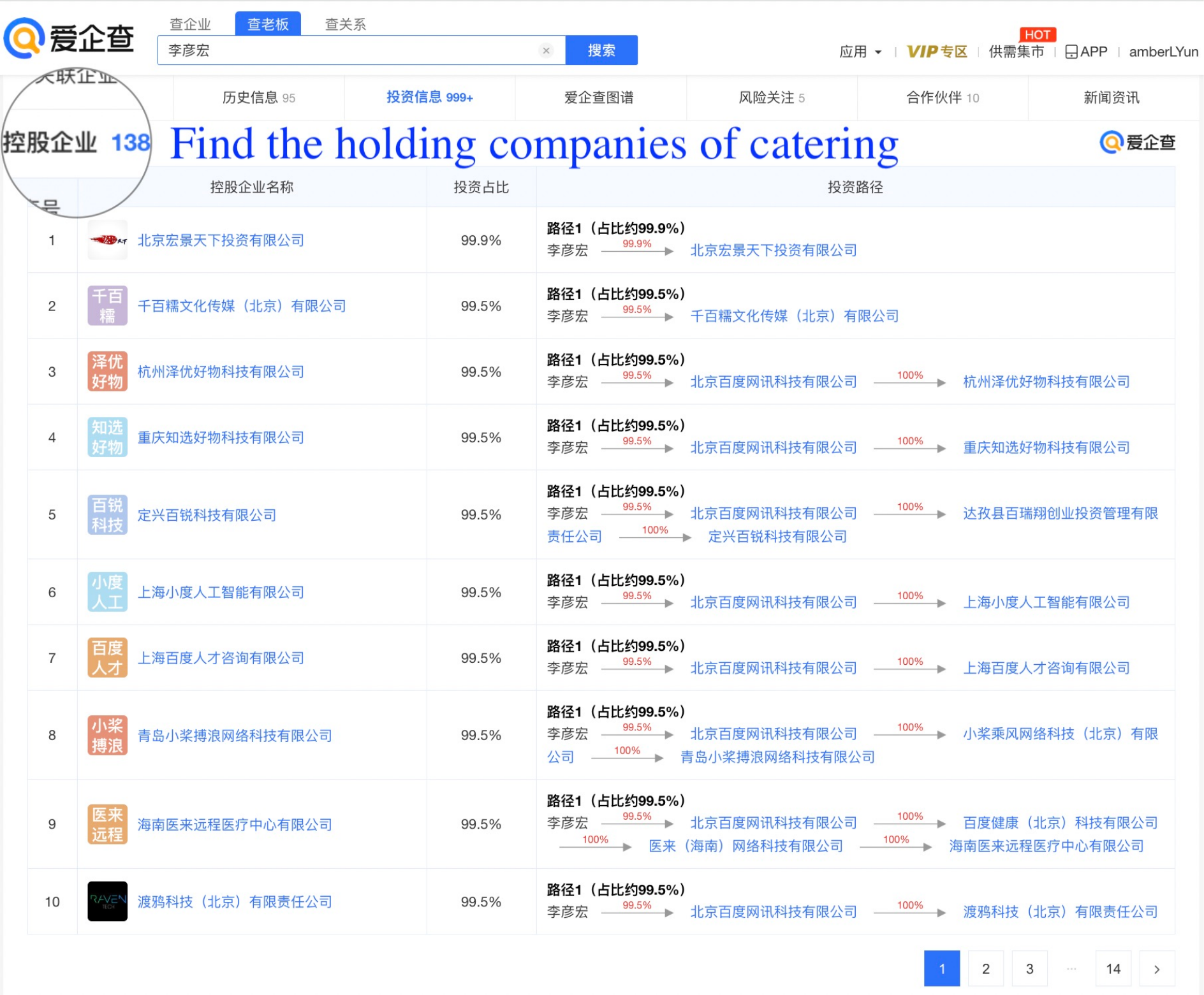}} \quad
    \subfloat[Chat-based Search and Exploration]{\includegraphics[width=0.42\textwidth]{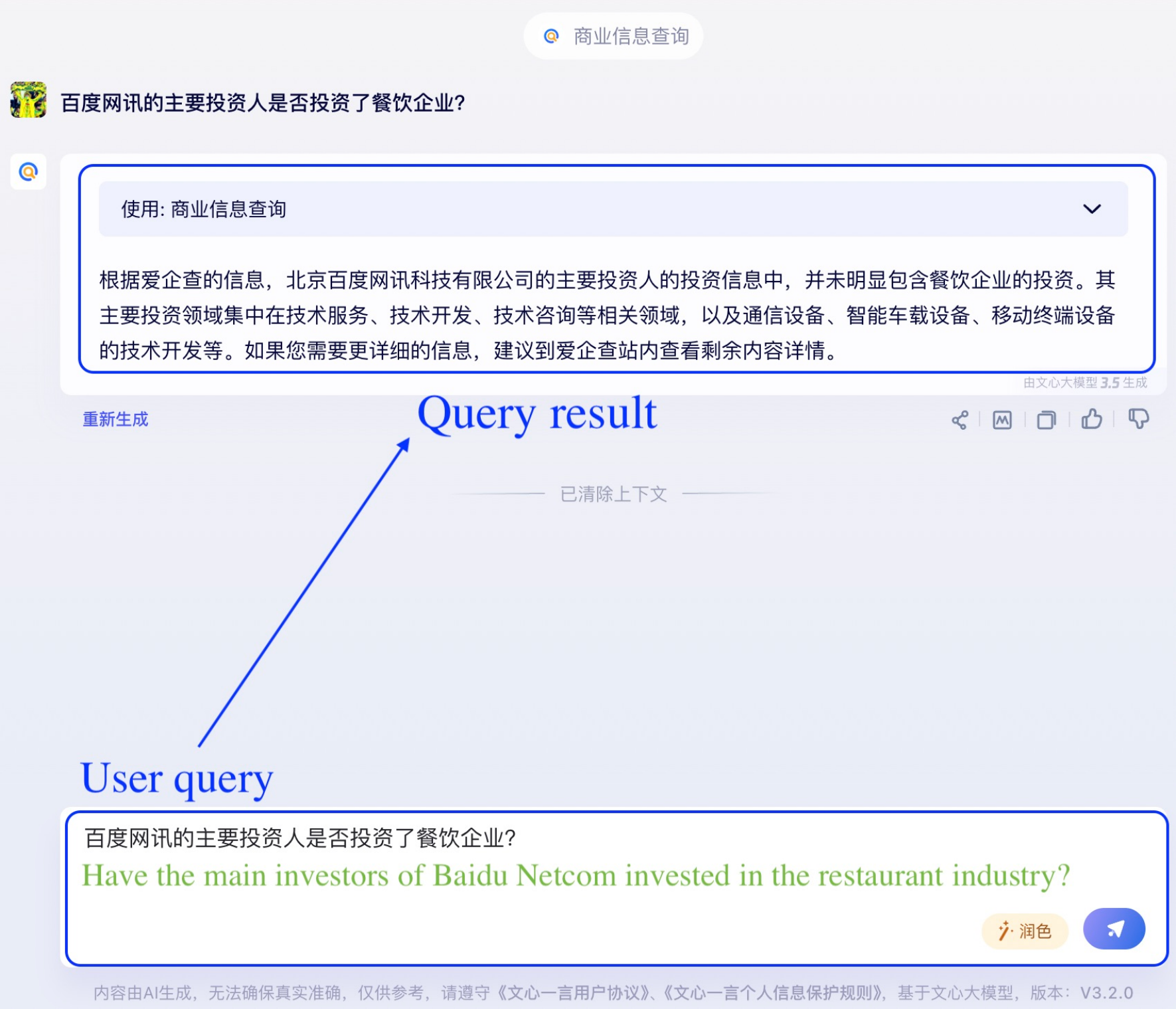}}\vspace{-2mm}
    \caption{Enterprise Information Search and Exploration: (a) conducting a manual search of the enterprise, (b) manually researching the major shareholders of the company, and (c) examining each shareholder to determine if they have invested in a catering company versus (d) chat-based search and summarization in one step.}\label{fig:search-explore}
    \vspace{-5mm}
\end{figure*}

Consider a scenario where a financial analyst is tasked with investigating whether the major shareholders of a company have made any investments in catering companies. The analyst must gather detailed information about these shareholders, including their investment patterns, related entities, and any associations with the catering industry.  As shown in Figure~\ref{fig:search-explore}, initially, the analyst would begin by querying the company's major shareholders, leading to an initial graph node that provides basic shareholder information. From there, the analyst needs to manually follow links to subgraphs that represent these shareholders' investment portfolios, affiliated business entities, and any connections to the catering sector. This manual exploration can be time-consuming and error-prone, as the analyst might overlook significant relationships or struggle to decipher intricate investment networks. This process may involve tracing back from the key shareholder node to various corporate or individual investment nodes, evaluating their stake percentages in catering companies, and interpreting the financial implications of each investment. The complexity increases when some investment nodes represent other corporations, requiring further layers of analysis.

Given these challenges, an effective solution would be an intelligent system that uses LLMs to understand natural language queries, explore knowledge graphs, perform complex queries, and summarize information. Such a system would not only save time but also significantly enhance the accuracy and comprehensiveness of the information retrieved, thereby transforming the user's search experience\cite{10387715,10417790}. In this work, we present \ours{}, deployed as a chatbot on Baidu Enterprise Search, that leverages the capabilities of LLMs to streamline the search, exploration, and summarization of enterprise information within a knowledge graph database containing nation-wide enterprise information. Such graph database, constructed using Apache TinkerPop, consists of hundred millions of nodes, ten billions of edges, hundred billions of node/edge attributes, and millions of subgraphs as company communities, reflecting the status of millions of companies, corporations, and organizations registered in China. 

To effectively initiate queries within the database, \ours{} incorporates a novel \textbf{data pre-processing pipeline} as follows. (1) \ours{} collects real-world web search queries related to enterprise information--such as company names, legal entities, and financial reports--from an general purpose search engines. (2) \ours{} selects representative queries as the \emph{seed dataset}, and makes developers carefully write search scripts for every query using Gremlin language for the TinkerPop-based enterprise graph database; (3) Subsequently, \ours{} builds a {vector database} comprising these representative queries and their scripts, achieved through meticulous data annotation and augmentation. By leveraging the {vector database} as prior, \ours{} secures the user experiences in retrieving and exploring the knowledge graph for enterprise information search through precise Gremlin search script generation with ICL.

\begin{figure*}
\centering
\includegraphics[width=0.85\linewidth]{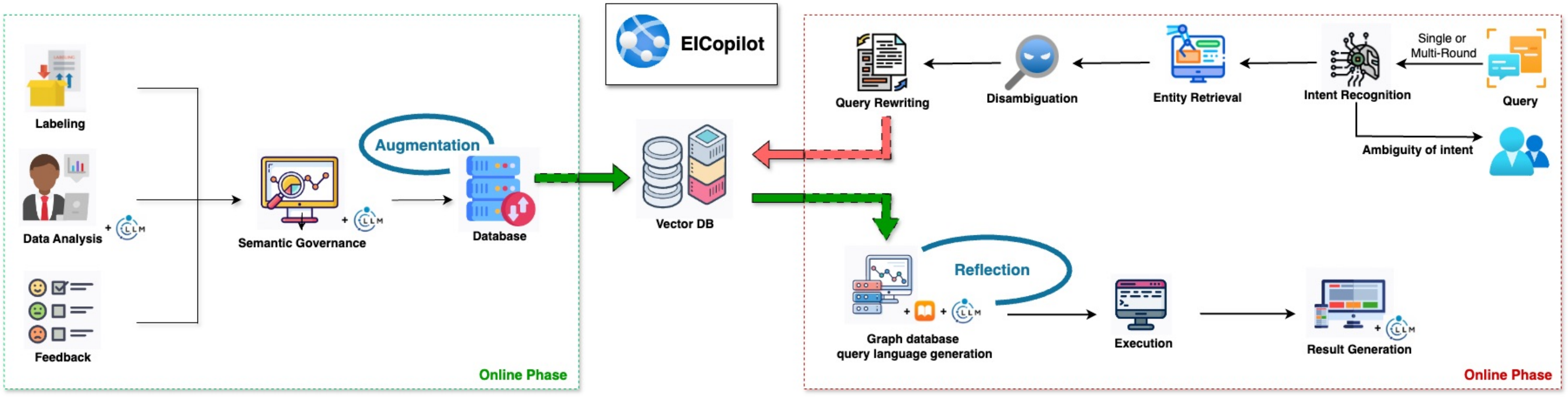}\vspace{-2mm}
\caption{The Overall framework of \ours{}.}
\label{fig:system_arch}
\vspace{-5mm}
\end{figure*}

Furthermore, when handling online requests, \ours{} provides precise query responses using a \textbf{comprehensive reasoning pipeline} based on Chain-of-Thought (CoT) and ICL. Specifically, our work reveals that searching the vector database often matches queries based on identical entity names, like company names, rather than similar search intent. Sometimes, we need examples (queries and scripts) with similar intent, such as those regarding a company's financial status or actual controller, which share similar syntax or logic in scripts. our work finds that masking entity names in queries enhances accuracy in intent matching. In this way, we propose a novel \textbf{query masking strategy} to improve ICL example matching and accurately interpret user intent, enhancing the precision of query script generation. \ours{} significantly outperforms baseline methods in speed and accuracy for data retrieval and interpretation, reducing the syntax error rate to as low as 10.00\% and achieving an execution correctness of up to 82.14\%. These results highlight the critical role of its components in enhancing query and summarization processes. 
While RAG provides a solid foundation, our work focuses on addressing the unique challenges of enterprise information search, such as domain-specific knowledge and complex user intents, where tailored solutions are essential. It represents a major advancement in exploring and utilizing large-scale knowledge graphs for enterprise information search. The technical contributions of this work are as follows.
\begin{itemize}
    \item This work explores technical problems in providing users a chat-based search and exploration experience for enterprise information, focusing particularly on the utilization of Gremlin over traditional query languages like SQL or GraphQL. By addressing the challenges inherent in automated script generation for Gremlin scripts, the research highlights the limitations of existing solutions and sets the foundation for more advanced methodologies tailored to handling complex graph database queries. This shift marks a significant step forward in improving the efficiency and accuracy of enterprise information retrieval processes.

    \item This work propsoed \ours{} with a robust framework that encompasses (i) a data pre-processing through representative queries collection, annotation, and augmentation, (2) a comprehensive reasoning pipeline that integrates CoT with ICL for enhanced query understanding, script generation, and RAG respectively, (3) a novel query masking strategy, which accurately identifies and matches user intent by masking specific entities within queries and corresponding database entries. \ours{} ensures that queries are not only understood in their complexity but are also executed with greater precision and relevance to user intent.

    \item The work includes extensive empirical analysis and experimentation based on real-world deployment with 5000$\sim$ DAU, demonstrating the capability of \ours{} to outperform baseline methods in data retrieval speed and accuracy across multiple scenarios. By evaluating various components of the agent-based methodology, the study provides valuable insights into their individual contributions and the overall effectiveness of the framework. This comprehensive analysis underscores the advancements \ours{} introduces in the field, marking it as a transformative tool for querying and summarizing complex data insights within large-scale knowledge graphs.

\end{itemize}

\section{Framework Design}
As illustrated in Figure~\ref{fig:system_arch}, \ours{} framework presents a sophisticated dual architecture for processing natural language queries directed towards enterprise information graph databases. During the offline phase, emphasis is placed on the preparation and enrichment of a robust data foundation to support subsequent online operations effectively. The construction of an enriched seed data repository, complemented by efficient data augmentation strategies, further enhances the system's capacity to generate precise query responses. Transitioning to the online phase, the intent understanding and decision-making modules of the \ours{} framework leverage the intricate capabilities of LLMs to interpret various user queries, accommodating non-standard query phrases. The system's capabilities in ICL and CoT further streamline the process, facilitating efficient and relevant information retrieval.

The integration of graph data query statement generation with contextual cues, coupled with the supervision provided by the reflection module, ensures the generation of accurate and logically consistent graph database information retrieval. These functionalities, along with the fundamental components utilized by the framework, attest to the precise, reliable, and secure operation of \ours{}. The resultant presentation constitutes the ultimate point of interaction between users and \ours{}, wherein complex structured data is transformed into textual responses, significantly enhancing the user experience. Subsequent sections will progressively delve into each phase to elucidate the detailed design principles.

\subsection{Offline Phase}
The offline phase is dedicated to preparing high-quality data that will be utilized during the online phase. This preparatory work involves four key activities: Schema Semantic Governance, Seed Data Construction, Data Augmentation, and Masked Question Similarity Selection. Once organized and refined, these data sets are embedded into an in-memory vector database, which in turn equips the online phase with the necessary contextual data to address specific user queries.

\subsubsection{Schema Semantic Governance}
Schema information is crucial for data analysis tasks, such as graph database query language writing, similar to the necessity for humans to comprehend complex graph databases. The \ours{}, by interpreting schema details, generates graph database query statements that align with human intent. For fields with enumerated values, it is essential to provide a comprehensive description of the values stored in the database and their interpretations. In cases involving complex data types like structures and maps, it is critical to elucidate key properties and their meanings. These detailed metadata facilitate an understanding of the intricate relationships between attributes, thereby enabling accurate response to queries for data retrieval. Additionally, annotations should be applied to graph database query languages to ensure the generation of syntax that meets the specific requirements of the target system.

\subsubsection{Seed Data Construction}
Current LLMs have limited \\ autonomous capabilities in generating query languages for knowledge graphs. By leveraging the ICL capabilities, we aim to enhance the stability and accuracy of knowledge graph query generation by integrating dynamic query pairs relevant to the user's current problem. There are three strategies for constructing these seed data pairs. The first involves embedding query pairs commonly found in enterprise workflows, especially since some queries frequently involve business-specific metrics. Without the aid of few-shot learning or fine-tuning mechanisms, it is unrealistic to expect LLMs to autonomously generate user-specific knowledge graph queries. The second approach utilizes Graph2NL technology, which supports a cold start process without necessitating user-supplied knowledge graph query pairs. This is particularly advantageous given the extensive data available in corporate data warehouses. In this scenario, natural language queries are generated using knowledge graph and schema information, thereby establishing knowledge graph query pairs. The final method involves creating knowledge graph query pairs based on user feedback in cases where accurate results are not produced. In such instances, human intervention becomes essential for correcting inaccuracies.

\subsubsection{Masked Question Similarity Selection}

To address the shortcomings in autonomous knowledge graph query generation capabilities, we adopted the ICL method. By dynamically integrating query pairs relevant to the user's current query, we enhanced the model's ability to generate stable and precise query statements. Specifically, this method improves the model's query generation capability by dynamically generating contextually relevant query pairs.

To prevent key information differences in user queries from decreasing the relevance of information retrieval and intent recognition, we standardized the usage of  query masking strategy. This standardization significantly increased the recall and accuracy of similar instances in the sample repository, thereby enhancing the overall effectiveness of the query generation mechanism.

In implementation, we focused on the following aspects:

\begin{enumerate}
    \item \textbf{Query Pair Construction}: As shown in section 3.1.2, We built a sample repository, collecting diverse query pairs relevant to potential user queries. Each query pair contains a question and a corresponding query statement. By carefully selecting example pairs as context, we significantly improved the accuracy and relevance of generated query statements.

    \item \textbf{Offline Data Population}: During the preprocessing phase, we standardized corporate entities to ensure consistent representation of the same entities across different queries. This standardization allows the model to more accurately identify and generate queries matching user intent.

    \item \textbf{Multi-turn Dialogue Capability}: The ICL strategy enables the model to gradually clarify and accurately identify the user's query intent in multi-turn dialogues. This capability allows users to ask complex questions naturally and ensures the model continues to understand and generate accurate queries.

    \item \textbf{Dynamic Updating}: As shown in section 3.1.4, To maintain high precision in query generation, we continuously update the example knowledge repository to reflect new user query intents and dynamic changes in corporate data.
\end{enumerate}

Through the implementation of these strategies, our knowledge graph query generation mechanism has achieved significant improvements. The accuracy and stability of query statements have both improved, and the model can better understand the user's complex query intent. The ICL approach provides us with an effective means for large language models to autonomously adapt to different query scenarios and generate high-quality knowledge graph queries.

\subsubsection{Analysis System}
The system aims to leverage a substantial amount of real-world user data to enhance the richness of the vector database. We have designed an offline analysis architecture that exports query execution records and categorize failed tasks and regenerate query statements. Subsequently, the regenerated queries are manually reviewed, and successful user query-statement pairs are imported into the vector database, thereby improving the quality and accuracy of the knowledge base.

The core functionality of the offline analysis architecture lies in utilizing exported query execution records to analyze and identify the reasons behind query failures. The analysis framework is capable of correcting issues by categorizing failed tasks and generating new query statements. Specifically, when a query execution fails, the system exports the relevant execution record, identifies the type of failure, and generates alternative query statements that better align with user needs based on task context and query intent, ultimately increasing the query success rate.

After manual review, the regenerated query statements that successfully execute are paired with the user questions and imported into the vector database, thereby improving the quality of the knowledge base. This process ensures continuous database optimization and provides more accurate and comprehensive query results. By constantly refining the query execution records and regeneration strategies, the system gradually accumulates a large number of effective user query-statement pairs, providing richer references for future queries.

\subsection{Online Phase}
The system provides a refined user experience through an online architecture consisting of a series of modules driven by a Named Entity Recognition (NER) model, an Entity Retrieval model based on Natural Language Processing Customization (NLPC), and a LLM. By recognizing user intentions and generating precise Knowledge Graph Query Languages, the system first disambiguates user queries to link them with existing entities or intents. It then connects these queries with the structure of the knowledge graph, specifying the knowledge graph query statements based on the user's inquiry and similar examples. An introspection module enhances accuracy by correcting errors in the query statements. Subsequently, the knowledge graph executes these queries and processes the results. The result generation module transforms these outcomes into text-based analysis, minimizing visual complexity while still providing comprehensive insights. Overall, the online phase is a seamless sequence that ensures accurate and user-aligned outputs from query interpretation to execution.

\subsubsection{Intent Understanding and Decision-Making}
In the real-world application, user queries are diverse, encompassing content beyond specified parameters, topics outside the domain scope, incomplete queries, and questions that challenge continuity. To address these complexities, a decision-making mechanism supported by \ours{} has been designed. This mechanism integrates three main components: comprehensive intent absorption, relevant knowledge retrieval, and response generation—all aimed at enhancing the precision of service provided to users. 
The system utilizes user inquiries, similar example queries, and predefined decision metrics to craft detailed prompts, optimizing the LLM’s responses. The comprehensive intent understanding component is designed to capture the essence of user needs, addressing gaps in multi-turn dialogues. Relevance assessment determines the pertinence of a query to the ongoing topic, thereby filtering out irrelevant or off-topic questions.

\subsubsection{Disambiguation}
The disambiguation process includes two steps: anaphora resolution and entity retrieval. For instance:

In the query, "What is the registered address of Baidu?" the term "Baidu" could refer to multiple corporate entities such as Baidu Netcom or Baidu Online. It is essential to determine the specific entity for the query before proceeding with data retrieval.
In the statement, "Who is their legal representative?" the term "their" is a pronoun that, within the context of a multi-turn dialogue, requires semantic understanding to resolve its reference based on the preceding conversation.
Initially, anaphora resolution is performed by considering the context of the historical dialogue to interpret the current user's query. The prompt for this process is as follows:






\begin{figure}[htbp]\vspace{-2mm}
    \centering
    \includegraphics[width=0.98\linewidth]{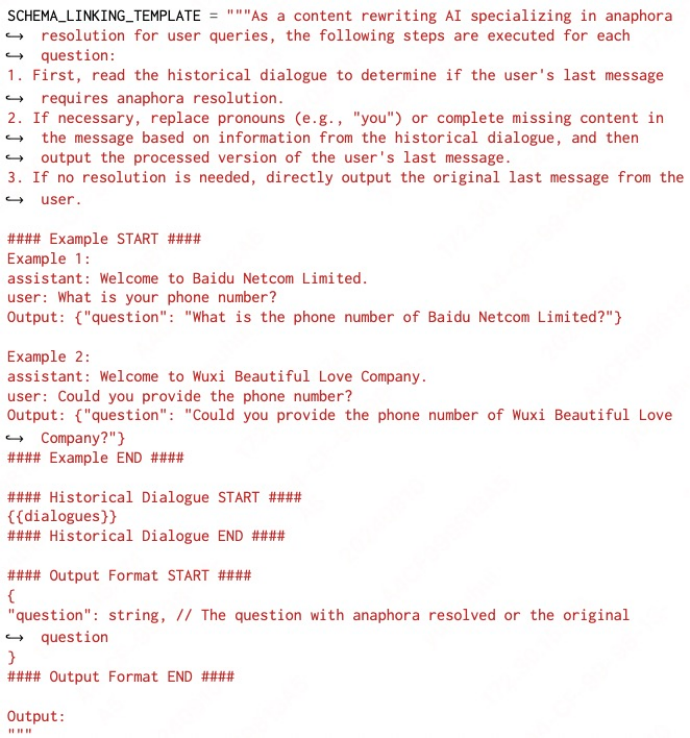}\vspace{-5mm}
    \caption{Schema linking prompt.}\vspace{-3mm}
    \label{fig:schema_linking}
\end{figure}








In the realm of LLMs, employing the aforementioned techniques can significantly reduce the occurrence of pronouns and incomplete sentences in queries. Following anaphora resolution, it is necessary to resolve ambiguities concerning corporate entities in the user’s query. This process relies on NER, NLPC, and an ElasticSearch-based system for corporate entity retrieval. Initially, an NER model segments the user’s query to extract corporate names. Then, a recall service that integrates NLPC and ElasticSearch is deployed for precise entity matching. If a match is found, the complete name and ID of the corporate entity are obtained, thus finalizing the disambiguation process. If the initial match attempt via NLPC and ElasticSearch fails, a fuzzy search is triggered. When multiple entities are detected, the user is prompted to confirm the specific entity. This ensures that the subject of the user's query is clearly identified and uniquely retrievable in the knowledge graph.

\subsubsection{Schema Linking}
In practical commercial environments, the challenge of managing large graph databases is often constrained by the context bandwidth of LLM prompts. This contextual limitation renders it impractical to integrate the comprehensive architectural details of all graph data tables into a single prompt. Moreover, incorporating excessive schema information could adversely affect the model's performance. As a solution, we introduce a Schema Linking Module, positioned prior to the construction of the graph database, to circumvent the complexity of excessive schema input.

Contemporary schema linking interventions typically revolve around the development of new models, which requires substantial resources and may face limitations in domain generality. To counteract this, our approach employs an initial multi-recurrence strategy to identify tables aligned with the user's query intent. Subsequently, the LLM is utilized to establish links between the user's question and the relevant graph database and fields.

We then derive the details of schema linking from the prescribed prompt, as illustrated in Figure \ref{fig:schema_linking}. This two-stage method facilitates efficient graph database query language generation for large tables within specific domains and simplifies the schema linking process, thereby enhancing the model's effectiveness and generalization potential.

\begin{figure}[htbp]
    \vspace{-3mm}
    \centering
    \includegraphics[width=0.98\linewidth]{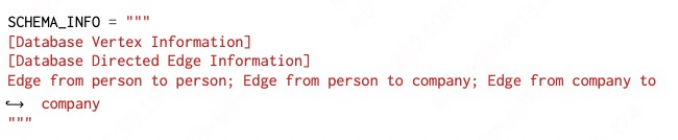}
    \vspace{-3mm}
    \caption{Schema information prompt.}\vspace{-3mm}
    \label{fig:schema_info}
\end{figure}

\subsubsection{Query Language Generation}
The module aims to accurately deconstruct user queries by obfuscating key information. The core process involves separating the user's query content from the intent and then leveraging a knowledge base in a vector database to match the intent. The specific workflow can be illustrated using the following example: "Who is the boss of Baidu Netcom Technology Company?" 

First, NER identifies the company entity in the query, and all company names are uniformly replaced, thus rewriting the question. Subsequently, the knowledge base in the vector database is used to perform similarity matching based on the rewritten question, selecting cases that most closely align with the user's intent and appending these cases to \ours{} to enhance its understanding of the query intent.

To further improve the agent's accuracy when generating graph database query language, the module converts the user's diverse query questions into domain-specific knowledge comprehensible to the agent. It clarifies certain lexical mappings, for instance, both the legal representative of a company and the actual controller can be mapped to "boss". In this way, the module deconstructs and abstracts the user's diverse query questions into more generic query intents and then matches them with the cases in the knowledge base to accurately comprehend the user's intent and generate precise responses.

\subsubsection{Reflection}
This module aims to significantly enhance the accuracy of graph database query languages to rectify erroneous query language expressions. Through thorough evaluation and comprehensive review, we meticulously verify the authenticity of graph database query commands, delving into the root causes of errors, including the validation of edges, edge directions, attributes, and the syntactic integrity of queries. The reflection mechanism is activated only after anomalies in the graph database query commands have been clearly identified.

The prompt configuration of this module is meticulously designed to not only specify error types but also include graph database queries, detailed graph structure information, and in-depth diagnostic assessments of discrepancies. This framework enables \ours{} to focus on correcting potential errors in graph database query commands, ensuring an accurate and targeted rectification process.

\subsubsection{Result Generation}
After retrieving the execution results from the database engine, the result generation module comes into play, constructing the final output for the web-based application. \ours{} is capable of providing recommendations for scenarios such as complaints and information inquiries based on the retrieved context and content, effectively guiding users for further actions. Additionally, the module integrates functions like product recommendations, corporate directories, and supply-demand marketplaces to address users' multifaceted needs. To enhance system response speed, as outlined in section 3.2.1, if a user's intent is identified as related to procurement, franchising, etc., a template-based response will be returned directly. For other intents, our system will provide a summarized response.

\section{EXPERIMENTAL EVALUATION}
In this section, we present the experimental setups, followed by a comparisons of our overall performance results. We then delve into ablation tests and conclude with an analysis of various case studies.  
\subsection{Experimental Setup}
We mainly introduce the experimental setups from the perspectives of LLM models and datasets used for evaluation.

\begin{figure}
    \centering
    \includegraphics[width=0.85\linewidth]{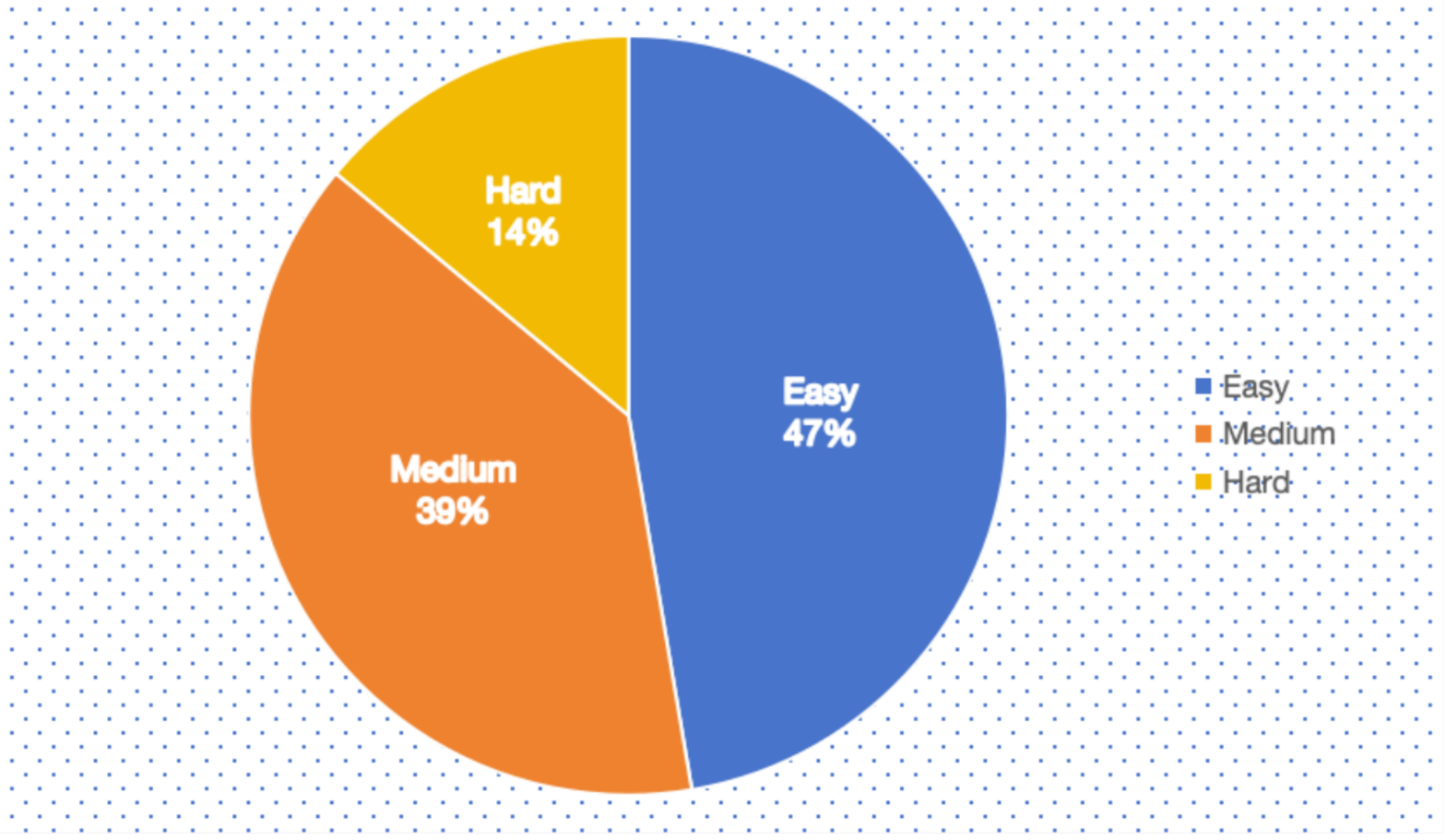}\vspace{-5mm}
    \caption{The difficulty of Real Traffic Dataset.}\vspace{-5mm}
    \label{fig:difficulty_new}
\end{figure}

\subsubsection{Dataset Construction.} 
Due to the lack of publicly available datasets, we obtained data from Baidu's internal data platform. Through rigorous processing, we constructed a test dataset consisting of 150 entries. Each entry comprises an input query paired with its corresponding graph database query statement. By decomposing the query into segments by splitting the string at each period, we determine the number of operational steps in the query traversal. The length of this list of steps provides an initial indication of complexity; typically, longer queries entail more intricate logic and higher resource consumption. To simplify the evaluation without losing generality, the complexity score based on traversal length is assigned as follows:
\begin{equation}
\text{Length Complexity Score} = 
\begin{cases}
1 & \text{if } \text{steps} < 5 \\
2 & \text{if } 5 \leq \text{steps} \leq 7 \\
3 & \text{else} \\
\end{cases}
\end{equation}
E.g., the query \texttt{g.V().hasLabel('person').values('name')} contains 3 steps, resulting in a score of 1. Conversely, a query with 8 steps would receive a score of 3. Each traversal step is further evaluated based on the operators used, categorized as shown in Table \ref{tab:operator_complexity}. Basic operations like \texttt{has} score 1 point, while more complex operations like \texttt{repeat} score 3 points due to their higher complexity.
The final complexity score is the sum of the traversal length score and the points assigned to the operations. The final difficulty level is determined as follows:

\begin{table}
    \caption{Operator Complexity Scores in Gremlin Queries.}\vspace{-3mm}
    \scalebox{0.8}{\begin{tabular}{ccc}
        \hline
        \textbf{Easy} & \textbf{Moderate} & \textbf{Complex} \\
        \textbf{(Basic Operations)} & \textbf{(Simple Aggregation)} & \textbf{(Advanced Operation)} \\
        \hline
        \texttt{has}       & \texttt{groupCount} & \texttt{repeat} \\
        \texttt{out}       & \texttt{fold}       & \texttt{times} \\
        \texttt{in}        & \texttt{select}     & \texttt{where} \\
        \texttt{values}    & \texttt{order}      & \texttt{path} \\
        \texttt{by}        & \texttt{dedup}      & \texttt{choose} \\
        \texttt{label}     & \texttt{count}      & \texttt{coalesce} \\
        \texttt{id}        & \texttt{sum}        & \texttt{union} \\
        \texttt{V()}       & \texttt{min}        & \texttt{project} \\
        \texttt{E()}       & \texttt{max}        & \texttt{branch} \\
                          & \texttt{mean}       & \texttt{match} \\
        \hline
        \textbf{1 Point}   & \textbf{2 Points}   & \textbf{3 Points} \\
        \hline
    \end{tabular}}
    \label{tab:operator_complexity}
\vspace{-3mm}
\end{table}

\begin{equation}
\text{Query Complexity} = 
\begin{cases}
\text{Simple} & \text{if } \text{Scores} \leq 4 \\
\text{Moderate} & \text{if } 5 \leq \text{Scores} \leq 7 \\
\text{Complex} & \text{else} \\
\end{cases}
\end{equation}
Consider the query \texttt{g.V().out('knows').groupCount().by('name')}. With 5 traversal steps, it receives a length complexity score of 2. The \texttt{out} operator adds 1 point, while \texttt{groupCount} and \texttt{by} each contributes 2 points, leading to a total score of 7, thereby classifying the query as moderate in complexity. As illustrated in Figure \ref{fig:difficulty_new}, the dataset used for evaluation is diverse and challenging.

\subsubsection{Evaluation metrics}
To assess the performance of the proposed \(\ours{}\) method, we consider two key perspectives: the syntax errors of the Gremlin scripts generated within the \ours{} and their execution correctness. The detailed definitions for each metric are as follows:
\begin{itemize}[leftmargin=*,noitemsep,nolistsep]
    \item \textbf{Syntax Error Rate}, defined as the percentage of predicted Gremlin scripts that are free of syntactic errors. It can be computed by $[1 - \frac{1}{N} \sum_{i=1}^{N}\vmathbb{1}(R_i)]$, where $\vmathbb{1}(\cdot)$ is an indicator function, which can be represented as $\vmathbb{1}(R) = \left\{\begin{matrix} 1, & \text{execution success}\\ 0, & \text{execution failed}\end{matrix} \right.$.

    \item \textbf{Execution Correctness}, defined as the proportion of queries rated for their effectiveness in fulfilling user requirements, whether directly or indirectly. This metric is derived from expert evaluations of generated Gremlin scripts, assessing factors like alignment with user intent and overall script reliability. Scripts that fully satisfy user needs receive a score of 1, those offering partial or potential assistance are scored 0.5, and those deemed irrelevant to user requirements receive a score of 0.
\end{itemize}

\subsubsection{LLM Models}
In this experiment, we compared the performance of \ours{} on top of three models as follows:
\begin{itemize}
    \item \textbf{ErnieBot} is a close-source, full-size LLM developed by Baidu, designed for understanding and generating human-like text across various applications, eg., code generation.

    \item \textbf{ErnieBot-Speed} is the lite and fast version of ErnieBot with fewer trainable parameters. This model supports SFT for application customization.
    
    \item \textbf{Llama3-8b} is an open-source LLM with 8 billion parameters. It supports SFT for applications.
\end{itemize}
To fine-tune ErnieBot-Speed and Llama3-8b, we collect and annotate a dataset consisting of 418 manually selected Gremlin query pairs that cover a wide range of Gremlin syntax. The data is divided into training and validation sets in an 8:2 ratio. During the fine-tuning process, we use a full SFT approach based on training samples and perform the parameter tuning with the validation set.

Notably, the ErnieBot and Llama series are utilized as foundational LLMs within our system, specifically for the prompt-tuning process of \ours{}. Although prompt-tuning is not our primary contribution, we emphasize system design and agentic workflows for chat-based enterprise information searches using knowledge graphs. Consequently, we limited evaluations to these models for consistency, without asserting their superiority in all Gremlin generation tasks. While other LLMs may achieve superior performance in these tasks through pre-training or fine-tuning, our contribution to system design can effectively complement such advancements. 

\subsubsection{Baselines and Configurations}
When an evaluation query serves as the target for the generation of Gremlin scripts, representative queries are used as potential examples for ICL. The \ours{} framework masks both the evaluating and representative queries, constructs a vector database, and executes similarity-based matching utilizing the following four strategies:
\begin{itemize}
    \item \textbf{\em Raw Matching (Raw Match)}: This strategy employs LLMs to extract vectors for both the evaluating and representative queries, performing similarity-based matching between the evaluating query and every representative query.
    
    \item \textbf{\em Representative Query Entity Masking (Rep. Mask)}: Here, the entities in each representative query are masked before performing vector-based matching.
    
    \item \textbf{\em Representative Query Entity Masking (Eval. Mask)}: Here, the entities in each representative query are masked before performing vector-based matching.
    
    \item \textbf{\em Full Entity Masking (Full Mask)}: This approach involves masking entities in both the evaluating and representative queries, followed by similarity-based matching of the vectors. This strategy is used in the production of \ours{}.
\end{itemize}
The top-3 and top-5 matched representative queries serve as few-shot examples for ICL in our experiments, respectively. Notably, the vectors of all representative query are pre-extracted offline to build the vector database.

\begin{table}
    \caption{Syntax Error Rate of Gremlin scripts generated by Zero-shot and ICL-based approaches. EB: ErnieBot (Pre-trained), EBS: ErnieBot-Speed (Pre-trained), EBS (SFT): ErnieBot-Speed (Fine-tuned), Llama: Llama3-8b (Pre-trained), Llama (SFT): Llama3-8b (Fine-tuned)}\vspace{-3mm}
    \centering
    \begin{tabular}{lcccccc}
        \hline
        Strategies& EB & EBS & EBS (SFT)  & Llama & Llama (SFT) \\        \hline
        \multicolumn{6}{c}{Zero-Shot}\\\hline
        & 15.33\% & 22.67\% & 16.67\% & 52.67\% & 30.00\% \\
         \hline
         \multicolumn{6}{c}{\ours{} + top-3 matched representative queries}\\\hline
         \emph{Raw Match} & 17.33\% & 25.33\% & 12.00\% & 32.67\% & 12.67\% \\
        \hline
         \emph{Eval. Mask} & 38.67\% & 48.67\% & 10.00\% & 47.33\% & 20.00\% \\
        \hline
         \emph{Rep. Mask} & 17.33\% & 27.33\% & 12.67\% & 27.33\% & 11.33\% \\
        \hline
          \emph{Full. Mask} & 20.00\% & 27.33\% & 14.00\% & 30.00\% & 13.33\% \\\hline
         \multicolumn{6}{c}{\ours{} + top-5 matched representative queries}\\\hline
         \emph{Raw Match} & 17.33\% & 25.33\% & 7.33\% & 30.00\% & 12.00\% \\
        \hline
         \emph{Eval. Mask} & 27.33\% & 40.67\% & 7.33\% & 52.00\% & 6.00\% \\
        \hline
         \emph{Rep. Mask} & 14.00\% & 26.00\% & 2.67\% & 11.33\% & 4.67\% \\
        \hline
        \emph{Full Mask} & \textbf{10.00\%} & 30.67\% & \textbf{2.00\%} & 12.00\% & 8.00\% \\ \hline
    \end{tabular}
    \label{tab:llm_result_1}
    \vspace{-5mm}
\end{table}

\begin{table}
    \caption{Execution Correctness of Gremlin scripts generated by Zero-shot and ICL-based approaches. EB: ErnieBot (Pre-trained), EBS: ErnieBot-Speed (Pre-trained), EBS (SFT): ErnieBot-Speed (Fine-tuned), Llama: Llama3-8b (Pre-trained), Llama (SFT): Llama3-8b (Fine-tuned)}\vspace{-3mm}
    \centering
    \begin{tabular}{lcccccc}
        \hline
        Strategies& EB & EBS & EBS (SFT)  & Llama & Llama (SFT) \\        \hline
        \multicolumn{6}{c}{Zero-Shot}\\\hline
        & 41.00\% & 17.33\% & 37.67\% & 17.66\% & 36.67\% \\
        \hline
         \multicolumn{6}{c}{\ours{} + top-3 matched representative queries}\\\hline
         \emph{Raw Match} & 45.33\% & 33.67\% & 42.33\% & 32.00\% & 43.33\% \\
        \hline
         \emph{Eval. Mask} & 39.67\% & 25.00\% & 56.33\% & 26.00\% & 28.00\% \\
        \hline
         \emph{Rep. Mask} & 49.67\% & 33.33\% & 48.67\% & 37.33\% & 49.33\% \\ \hline
        \emph{Full. Mask} & 51.67\% & 42.67\% & 51.33\% & 42.00\% & 51.33\% \\ \hline
         \multicolumn{6}{c}{\ours{} + top-5 matched representative queries}\\\hline
        \emph{Raw Match} & 57.33\% & 55.00\% & 57.67\% & 53.00\% & 58.00\% \\
        \hline
         \emph{Eval. Mask} & 50.00\% & 41.33\% & 61.67\% & 33.00\% & 70.67\% \\
        \hline
         \emph{Rep. Mask} & 76.79\% & 61.61\% & 81.25\% & 65.18\% & 74.11\% \\
        \hline
        \emph{Full Mask} & \textbf{82.14\%} & 65.18\% & \textbf{83.93\%} & 69.64\% & 79.46\% \\ \hline
    \end{tabular}
    \label{tab:llm_result_2}
    \vspace{-5mm}
\end{table}

\subsection{Performance Comparisons}
In this section, we compare the performance of \ours{} in various settings compared to the zero-shot approach. We collect a testing dataset with 150 queries from the real-world traffics of Aiqicha and annotate these queries with Gremlin scripts. Note that there is no overlap between the testing dataset for evaluation and the training/validation datasets for SFT.


Table \ref{tab:llm_result_1} reveals that \ours{} under various models improves syntax error rate, with the \emph{Full Mask} variant achieving an impressive 10.00\% and 2.00\%, indicating superior query quality. Other variants, such as \emph{Raw Match} and \emph{Rep. Mask}, also show improvements over the zero-shot baseline, which averages 19.33\%. In terms of execution correctness, shown in Table \ref{tab:llm_result_2}, \ours{} with \emph{Full Mask} excels with top scores up to 83.93\%, underscoring its practical usability. The \emph{Rep. Mask} variant performs strongly across configurations, reaching 81.25\%, both outperforming the zero-shot approach, which ranges between 17.33\% to 41.00\%. 
In general, \ours{} consistently improves both syntax quality and execution correctness across models, with \emph{Full Mask} providing the most notable improvements, thus achieving high query language quality and usability.

\subsection{Ablation Analysis}
We here analyze the four masking and representative query matching strategies, assessing their impact on syntax error rate and execution correctness as shown in Tables \ref{tab:llm_result_1} and \ref{tab:llm_result_2}.

Without masking any entities in both the evaluating and representative queries, \emph{Raw Match} still offers moderate gains, reducing syntax errors to 17.33\%-25.33\% and improving execution correctness to 53.00\%-58.00\%. Later, we can observe that masking either the evaluating query or the representative queries could bring performance gain: \emph{Eval. Mask} reduces syntax errors with results up to 40.67\% and correctness of 41.33\%-70.67\%. In contrast, \emph{Rep. Mask} also shows notable improvements, with syntax errors decreasing to 14.00\% and correctness rising to 61.61\%-81.25\%. 

By combining \emph{Eval. Mask} and \emph{Rep. Mask} strategies, \emph{Full Mask} excels, minimizing syntax errors to 10.00\%-2.00\% and achieving the highest correctness rates of 82.14\%-83.93\%. This strategy emerges as the most effective, consistently reducing errors and enhancing correctness, making it the best \(\ours{}\) configuration.

\subsection{Case Study}
We compare the alike querues matching results of \ours{} in the previous three strategies with a real-world online query: ``\textbf{\em Who are the executives of Binzhou Binxin Entertainment Network Technology Co., Ltd.?}'' Here, ``Binzhou'' refers to a small town in China, and ``Binxin'' is a brand name. As stated in Section 2.2.1, \ours{} would first decompose the online query to extract an entity -- ``\textbf{\em Binzhou Kaixin Entertainment Network Technology Co., Ltd.}'' and identify the search intent -- ``\textbf{\em company executives}''. 
\begin{figure}
    \centering
    {\fontfamily{qag}\selectfont\small
    \begin{enumerate}
        \item Which province is Binzhou Kaixin Entertainment Network Technology Co., Ltd. located in?
        \item Resumes of the management of Guangzhou L'Oréal Baiku Network Technology Co., Ltd.
        \item What is the phone number of the boss of Yirong (Henan) Information Technology Group Co., Ltd.?
        \item Who are the directors of Zhejiang Jiuzhou New Energy Technology Co., Ltd.?
        \item When was Huajing Entertainment Technology Co., Ltd. established?
        \end{enumerate}}\vspace{-3mm}
    \caption{Top-5 Matched Representative Queries using Raw matching}\vspace{-2mm}
    \label{fig:raw-match}
\end{figure}

\begin{figure}
    \centering
    {\fontfamily{qag}\selectfont\small
    \begin{enumerate}
    \item Resumes of Baidu's management
    \item Who are Baidu's investors?
    \item Who are Baidu's shareholders with more than 5\% ownership?
    \item What are the related companies of Baidu's legal representative?
    \item Who are the directors of Baidu?
    \end{enumerate}}\vspace{-2mm}
    \caption{Top-5 Matched Representative Queries using Rep. Mask}\vspace{-3mm}
    \label{fig:rep-mask}
\end{figure}

\begin{figure}
    \centering
    {\fontfamily{qag}\selectfont\small
    \begin{enumerate}
    \item Baidu's leadership
    \item Resumes of Baidu's management
    \item Information about Baidu's boss
    \item Who are Baidu's main responsible persons?
    \item Who is Baidu's boss?
    \end{enumerate}}\vspace{-3mm}
    \caption{Top-5 Matched Representative Queries using Full Mask (\ours{})}\vspace{-2mm}
    \label{fig:full-mask}
\end{figure}

Figures~\ref{fig:raw-match},~\ref{fig:rep-mask}, and~\ref{fig:full-mask} illustrate the top ``alike queries'' matched with the online query by the three strategies. The \emph{Raw Matching} strategy, which extracts queries without masking, results in examples related to general details of the targeted entity, but does not address the specific search intent (ie, searching for business executives).The results of the \emph{Representative Query Entity Masking} and \emph{Full Entity Masking} strategies, while focused on a different company—Baidu—align closely with the search intent of the online query, such as finding business executives, bosses, or leadership of a company. This alignment offers more relevant and targeted examples for generating scripts that accurately match the original search intent. Among these, full entity masking proves the most effective, as it enhances focus on search intent by masking entities in both online and representative queries, allowing for better generalization and capturing broader query meanings. This approach consistently generates relevant and thematically consistent queries, improving the quality and applicability of script generation by aligning closely with the original search intent.

\section{Discussions on Related Works}
This section explores key areas of our methodology: Text2SQL, RAG~\cite{gao2023retrieval}, and Information Retrieval (IR) \cite{zhu2023large}, which enhance querying and summarization of complex enterprise data from graph databases. Traditional IR systems, typically reliant on keyword matching, struggle with synonymy, polysemy, and contextual gaps, necessitating manual intervention \cite{zhu2023large}. Leveraging large pre-trained language models promises improvements across user modeling, indexing, matching/ranking, evaluation, and user interaction components \cite{bubeck2023sparks}. ICL allows models to adapt during inference, while RAG augments LLMs with external databases to reduce hallucinations and improve accuracy \cite{ding2024survey, borgeaud2022improving}, integrating retrieval and generation by enhancing input with retrieved data \cite{lozano2023clinfo, poesia2022synchromesh}.


Our method overcomes traditional IR limitations by using LLMs alongside ICL and advanced masking strategies, boosting semantic comprehension and reducing manual efforts \cite{ding2024survey, brown2020language}. We also tackle schema discrepancies by generating Gremlin scripts and applying ICL, advancing our approach as a leading solution in enterprise information retrieval \cite{androutsopoulos1995natural, deng2022recent}. Furthermore, the evolution of natural language to SQL translation, highlighted by works like NL2SQL-RULE \cite{guo2019content} and RATSQL \cite{wang2019rat}, addresses challenges like schema integration and query phrasing \cite{deng2022recent}.

\section{Conclusion}
This paper introduces \ours{} for enterprise information search that leverages LLMs to enhance querying and summarization in large graph databases. Key innovations include automated Gremlin script generation and a novel masking strategy for precise intent recognition in ICL example matching. Empirical analysis shows that \ours{} far exceeds baseline methods in data retrieval and interpretation speed and accuracy, potentially revolutionizing large-scale knowledge graph exploration. Specifically, \ours{} reduces syntax errors to as low as 10.00\% and increases execution correctness to 83.93\%. These results demonstrate the superior efficacy of its automated Gremlin script generation and innovative query masking strategies over traditional methods. This research offers valuable insights into agent-based methodologies and facilitates broader industrial applications in the exploration and utilization of large-scale graph databases.


\bibliography{Article}
\bibliographystyle{ACM-Reference-Format}

\appendix
\section{Deployment}
\begin{figure}[htbp]
    \centering
    \includegraphics[width=0.80\linewidth]{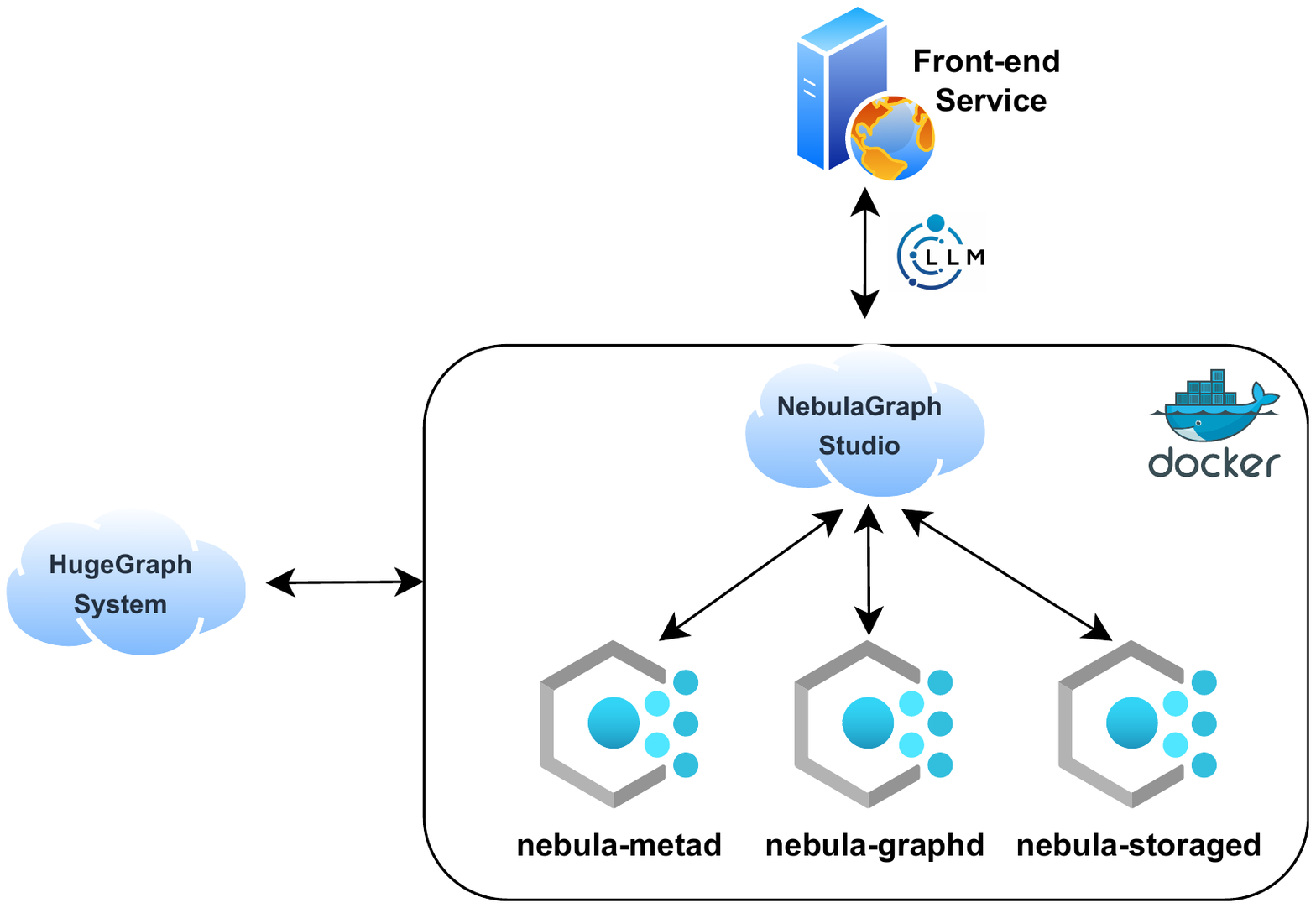}
    \caption{Graph database deployment architecture}
    \label{fig:graph-deployment-architecture}
\end{figure}

The deployment of our system, illustrated in Figure \ref{fig:graph-deployment-architecture}, employs a distributed architecture utilizing NebulaGraph and Docker. NebulaGraph is set up to function across multiple nodes to optimize data management. The HugeGraph system comprises hundreds of millions of nodes, with each node containing an average of 150 attribute fields. The architecture consists of several Docker containers, each containing components such as nebula-metad, nebula-graphd, and nebula-storaged, which enhances modularity and scalability. This setup facilitates the seamless management of graph data and supports efficient query processing. Furthermore, we use NebulaGraph Studio for data visualization and interaction. The system is meticulously configured to enhance the performance and reliability of graph operations. Particular attention is given to data distribution and replication across nodes to ensure robustness and improved access speeds.  

\begin{table*}
\centering
\small
\caption{End-to-end Experience: Examples of user input queries and Gremlin scripts generated by \ours{} for search on the graph database}
\begin{tabular}{|%
    >{\raggedright\arraybackslash}p{0.48\textwidth}|%
    >{\raggedright\arraybackslash}p{0.48\textwidth}|%
}
\hline
\textbf{User Input Query} & \textbf{Gremlin Script generated by \ours{}} \\
\hline
What is the corporate postal code of Guizhou Zhixun Tongchuang Technology Co., Ltd.? &
\begin{minipage}[t]{\linewidth}
\begin{minted}[breaklines, fontsize=\small]{python}
g.V().has('company','name','Guizhou Zhixun Tongchuang Technology Co.').values('postalCode')
\end{minted}
\end{minipage} \\
\hline
Could you provide the identity information of the legal representative of Linyi Juyun Trading Co., Ltd.? &
\begin{minipage}[t]{\linewidth}
\begin{minted}[breaklines, fontsize=\small]{python}
g.V().has('company', 'name', 'Linyi Juyun Trading Co., Ltd.').in('legalPerson').valueMap()
\end{minted}
\end{minipage} \\
\hline
Who are the executives of Reignwood FMCG Investment Management Co., Ltd.? &
\begin{minipage}[t]{\linewidth}
\begin{minted}[breaklines, breakanywhere, fontsize=\small]{python}
g.V().has('company','name','Reignwood FMCG Investment Management Co.,Ltd.').inE('serve').as('a').outV().as('b').project('name','position').by(select('b').values('name')).by(select('a').values('position'))
\end{minted}
\end{minipage} \\
\hline
\end{tabular}
\label{tab:query_gremlin_pairs}
\end{table*}

\section{Examples of Prompts and Outputs}
\subsection{Prompt: Gremlin Generation.}
\begin{figure}[htbp]
    \centering
    \includegraphics[width=0.90\linewidth]{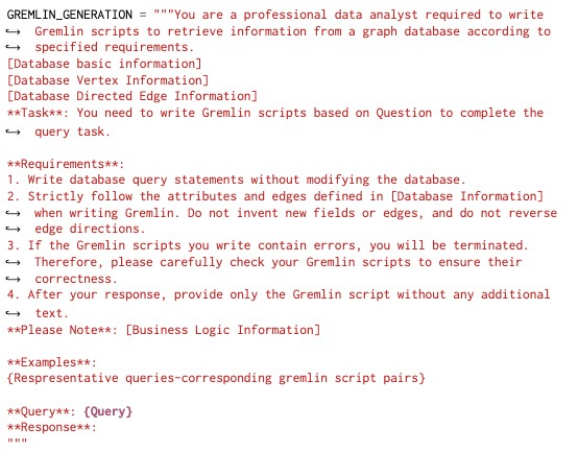}
    \caption{Gremlin generation prompt}
    \label{fig:gremlin-generation}
\end{figure}

The Gremlin generation prompt, as shown in Figure \ref{fig:gremlin-generation}, enhances accuracy and precision in script generation by integrating knowledge graph data and business logic. This comprehensive approach improves query reliability and expressiveness while maintaining data security and aligning with business needs. The content within \emph{[]} remains undisclosed due to its confidential nature related to business operations. Subsequently, we present several illustrative examples to demonstrate user inputs and the corresponding scripts generated by our Gremlin Generation prompts, as shown in Table \ref{tab:query_gremlin_pairs}.



\begin{table*}
\centering
\caption{Top-5 matched representative queries and their annotated gremlin scripts, for the user input query \emph{``What is the official website of World Kitchen (Shanghai) Co., Ltd.?''}, from the Vector Database}
\small
\begin{tabular}{|%
    >{\raggedright\arraybackslash}p{0.48\textwidth}|%
    >{\raggedright\arraybackslash}p{0.48\textwidth}|%
}
\hline
\textbf{Representative Query} & \textbf{Corresponding Gremlin Script} \\
\hline
What is Baidu’s contact number? &
\begin{minipage}[t]{\linewidth}
\begin{minted}[fontsize=\small, breaklines, breakanywhere]{python}
g.V().has('company','name','Baidu').values('phone')
\end{minted}
\end{minipage} \\
\hline
What is the latest news from Baidu? &
\begin{minipage}[t]{\linewidth}
\begin{minted}[fontsize=\small, breaklines, breakanywhere]{python}
g.V().has('company','name','Baidu').as('a').project('Company Information', 'Legal Person', 'Number of Overseas Investment Enterprises', 'Investor - Natural Person', 'Executive', 'Investor - Company', 'Ultimate Beneficiary - Natural Person', 'Ultimate Beneficiary - Company').by(valueMap('description','email','phone','operatingStatus','registrationAddress','salaryTreatment','registeredCapital','registeredCapitalCurrency','financingInformation')).by(select('a').in('legalPerson').values('name')).by(select('a').out('companyInvest').values('name').count()).by(select('a').in('personInvest').values('name').fold()).by(select('a').in('companyInvest').values('name').fold()).by(select('a').in('serve').limit(3).values('name').fold()).by(select('a').in('finalBeneficiaryPerson').values('name').limit(3).fold()).by(select('a').in('finalBeneficiaryCompany').limit(3).values('name').fold())
\end{minted}
\end{minipage} \\
\hline
What is Baidu? &
\begin{minipage}[t]{\linewidth}
\begin{minted}[fontsize=\small, breaklines, breakanywhere]{python}
g.V().has('company','name','Baidu').valueMap('name','description','industry','city','province','establishmentDate','listingDate','listingStatus','operatingStatus','email','phone','registrationAddress','website')
\end{minted}
\end{minipage} \\
\hline
Baidu &
\begin{minipage}[t]{\linewidth}
\begin{minted}[fontsize=\small, breaklines, breakanywhere]{python}
g.V().has('company','name','Baidu').as('a').project('Company Information','Legal Representative','Number of External Investment Enterprises','Investor - Individual','Executive','Investor - Corporate','Ultimate Beneficiary - Individual','Ultimate Beneficiary - Corporate').by(valueMap('description','email','phone','operatingStatus','registrationAddress','salaryTreatment','registeredCapital','registeredCapitalCurrency','financingInformation')).by(select('a').in('legalPerson').values('name')).by(select('a').out('companyInvest').values('name').count()).by(select('a').in('personInvest').values('name').fold()).by(select('a').in('companyInvest').values('name').fold()).by(select('a').in('serve').limit(3).values('name').fold()).by(select('a').in('finalBeneficiaryPerson').values('name').limit(3).fold()).by(select('a').in('finalBeneficiaryCompany').limit(3).values('name').fold())
\end{minted}
\end{minipage} \\
\hline
What is the official website address of Baidu? &
\begin{minipage}[t]{\linewidth}
\begin{minted}[fontsize=\small, breaklines, breakanywhere]{python}
g.V().has('company','name','Baidu').values('website')
\end{minted}
\end{minipage} \\
\hline
\end{tabular}
\label{tab:query_gremlin_pairs}
\end{table*}

\section{Scripts of Top Matched Representative Queries}
We provide the Top-5 representative query-Corresponding gremlin script pairs retrieved from the vector database, as shown in Table 5.

\end{document}